# Observer model optimization of a spectral mammography system

Erik Fredenberg,[a] Magnus Åslund,[b] Björn Cederström,[a] Mats Lundqvist[b] and Mats Danielsson[a]

[a]Department of Physics, Royal Institute of Technology, AlbaNova, 106 91, Stockholm, Sweden
[b]Sectra Mamea AB, Smidesvägen 2, 171 41 Solna, Sweden

**ABSTRACT**

Spectral imaging is a method in medical x-ray imaging to extract information about the object constituents by the material-specific energy dependence of x-ray attenuation. Contrast-enhanced spectral imaging has been thoroughly investigated, but unenhanced imaging may be more useful because it comes as a bonus to the conventional non-energy-resolved absorption image at screening; there is no additional radiation dose and no need for contrast medium. We have used a previously developed theoretical framework and system model that include quantum and anatomical noise to characterize the performance of a photon-counting spectral mammography system with two energy bins for unenhanced imaging. The theoretical framework was validated with synthesized images. Optimal combination of the energy-resolved images for detecting large unenhanced tumors corresponded closely, but not exactly, to minimization of the anatomical noise, which is commonly referred to as energy subtraction. In that case, an ideal-observer detectability index could be improved close to 50% compared to absorption imaging. Optimization with respect to the signal-to-quantum-noise ratio, commonly referred to as energy weighting, deteriorated detectability. For small microcalcifications or tumors on uniform backgrounds, however, energy subtraction was suboptimal whereas energy weighting provided a minute improvement. The performance was largely independent of beam quality, detector energy resolution, and bin count fraction. It is clear that inclusion of anatomical noise and imaging task in spectral optimization may yield completely different results than an analysis based solely on quantum noise.

## 1. INTRODUCTION

The energy dependence of x-ray attenuation is material specific because of (1) different dependence on the atomic number for the photo-electric and Compton cross sections ($\sigma \propto Z^4$ and $Z$ respectively), and (2) discontinuities in the photo-electric cross section at absorption edges. Spectral imaging is a method in medical x-ray imaging that takes advantage of the energy dependence to extract information about the object constituents.[1,2]

Phantom studies[3, ?, ?, 26, 38] and clinical trials[4] have proven contrast-enhanced spectral imaging to be a promising approach for enhanced tumor detectability. Injection of contrast agent is, however, probably not motivated for regular screening, and contrast-enhanced spectral imaging is expected to be an alternative mainly for diagnostic mammography at recalls. Enhancement of lesions without iodine uptake would be useful, since, in the case of electronic spectrum splitting, it comes as a bonus on top of the conventional absorption image with no additional dose to the patient. Spectral imaging could potentially increase detectability of obscured lesions, and discriminate between solid and cystic lesions already in the screening image, for instance.

Previous studies in this field have focused mainly on calcifications, predominantly with encouraging results,[5–10] although some are more moderate.[11] The main difficulty appears to be that amplified quantum noise in the subtracted image may reduce detectability of small details. For tumor imaging, a few clinical investigations have been presented.[12, 13] There are also phantom studies that indicate feasibility for soft tissue imaging,[14, 15] but

---

the minimal attenuation difference between glandular and tumorous tissue, which is likely the main challenge,[16] does not seem to be addressed.

In this work, we have investigated spectral imaging of tumors and microcalcifications without contrast agent, henceforth referred to as unenhanced imaging. There are at least three potential benefits of this approach compared to conventional non-energy resolved imaging, henceforth referred to as absorption imaging. (1) Energy weighting refers to optimization of the signal-to-quantum-noise ratio with respect to its energy dependence; photons at energies with larger agent-to-background contrast can be assigned a greater weight.[17,18] (2) Energy subtraction (or dual-energy subtraction) refers to optimization of the signal-to-background-noise ratio by minimization of the background clutter contrast. The contrast between any two materials (adipose and glandular tissue) in a weighted subtraction of images acquired at different mean energies can be reduced to zero, whereas all other materials to some degree remain visible.[3,4,6,15,19] (3) A third possible benefit of spectral imaging is quantification of the target, e.g. evaluation of microcalcification thickness.[10]

One way of obtaining spectral information is to use two or more input spectra. For imaging with clinical x-ray sources, this most often translates into several exposures with different beam qualities (different acceleration voltages, filtering, and anode materials).[3,4,6] Results of the dual-spectra approach are promising, but the examination may be lengthy with increased risk of motion blur and discomfort for the patient. This problem can be solved by a simultaneous exposure with different beam qualities,[20] or by using an energy sensitive sandwich detector.[8,9] For all of the above approaches, however, the effectiveness may be impaired due to overlap of the spectra, and a limited flexibility in choice of spectra and energy levels. In recent years, photon-counting silicon detectors with high intrinsic energy resolution, and, in principle, an unlimited number of energy levels (electronic spectrum-splitting) have been introduced as another option.[15,19]

An objective of the EU-funded HighReX project is to investigate the benefits of spectral imaging in mammography.[21] The systems used in the HighReX project are based on the Sectra MicroDose Mammography system (Sectra Mamea AB, Solna, Sweden), which is a scanning multi-slit full-field digital mammography system with a photon-counting silicon strip detector.[22,23] One advantage of this geometry in a spectral imaging context is efficient intrinsic scatter rejection.[24]

We have investigated a prototype detector and system[25] developed within the HighReX project for unenhanced spectral imaging. A semi-empirical cascaded system model and a framework for system characterization have been presented previously and is used also in the present study.[26] We have used an ideal-observer detectability index, which includes quantum and anatomical noise, as a figure of merit to investigate feasibility and for optimization.

## 2. MATERIAL AND METHODS

### 2.1. Background

#### 2.1.1. System Description

Figure 1 (Left) shows a photograph and schematic of the multi-slit system. The x-ray beam is collimated to a fan beam matching the pre-collimator. The pre-collimator transforms the beam to several equidistant line beams. Beneath the breast support there is a detector box containing a post-breast collimator and the x-ray detector. The detector is comprised of several lines of Si-strip detectors matching the line beams exiting the breast. The fan beam, pre-collimator, post-breast collimator and detector are scanned together laterally across the breast to obtain a full field image.

The detector that was used for measurements and simulations of spectral imaging is a prototype photon-counting detector, developed within the HighReX project and mounted on a Sectra MicroDose Mammography unit. A previous publication provides an investigation of the detector energy response, which is a prerequisite to accurately model spectral imaging.[25] Figure 1 (Right) shows a schematic of the detector. A bias voltage is applied over the detector material, so that when a photon interacts, charge is released and drifts as electron-hole pairs towards the anode and cathode respectively. Each strip is wire bonded to a preamplifier and shaper, which are fast enough to allow for single photon-counting. The preamplifier and shaper collect the charge and convert it to a pulse with a height that is proportional to the charge and thus to the energy of the impinging photon.

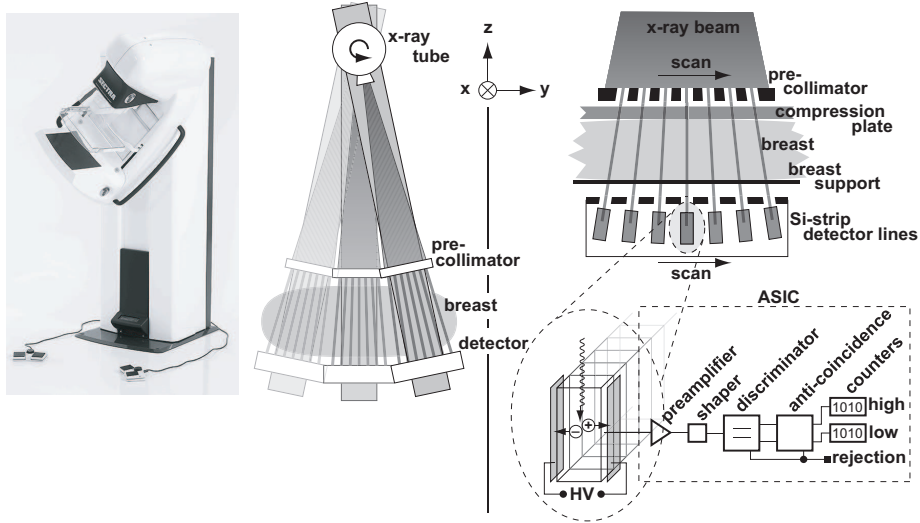

**Figure 1. Left:** Photograph and schematic of the Sectra MicroDose Mammography system. **Right:** The image receptor and electronics.

Pulses below a few keV are regarded as noise and are rejected by a low-energy threshold in a discriminator. All remaining pulses are sorted into two energy bins by an additional high-energy threshold, and registered by two counters. A preamplifier, shaper, and discriminator with counters are referred to as a channel, and all channels are implemented in an application specific integrated circuit (ASIC). Anti-coincidence (AC) logic in the ASIC detects double counting from charge sharing by a simultaneous detection in two adjacent channels, and the event is registered only once in the high-energy bin of the channel with the largest signal. Spatial resolution and image noise is thus improved, but all energy information of charge-shared photons is lost.

### 2.1.2. Observer modeling

For task-specific system performance, we can define an ideal-observer detectability index

$$d'^2 = 2\pi \int_{\text{Ny}} \text{GNEQ}(\omega) \times C^2 \times F^2(\omega) \times \omega \, d\omega, \quad (1)$$

where the integral is taken over the Nyqvist region and $\omega$ is the spatial frequency in the radial direction. Polar coordinates have been used for notational convenience, but all calculations were done in cartesian coordinates where appropriate, i.e. rotational symmetry was not assumed in general. $C = \Delta s/\langle I \rangle$ is the target-to-background contrast in the image for signal difference $\Delta s = \langle |I_{\text{background}} - I_{\text{target}}| \rangle$, where the angle brackets denote the expectation value and $I$ is the image signal per unit area. $F$ is the signal template, which integrates to the area of the target for unit contrast. GNEQ is the generalized noise-equivalent quanta that includes detector and anatomical noise according to Richard and Siewerdsen.[27–30] It is a reasonable extension of the standard noise equivalent quanta (NEQ) that measures the detector noise performance, because the dominant source of distraction for many imaging tasks in mammography is the variability of the anatomical background, which is referred to as anatomical or structured noise.[31, 32] For a quantum-limited system,

$$\text{GNEQ}(\omega) = \frac{\langle I \rangle^2 T^2(\omega)}{S_Q(\omega) + S_A(\omega)}, \quad (2)$$

where $T$ is the modulation transfer function (MTF) of the system, and $S_Q$ and $S_A$ are the power spectra (NPS) of quantum and anatomical noise respectively.

### 2.1.3. A theoretical framework for spectral imaging

A framework to characterize the performance of the multi-slit system for spectral mammography has been presented previously,[26] and is summarized below.

As was mentioned above, two spectral optimization schemes that appear in the literature are energy weighting and energy subtraction. Somewhat simplified, energy weighting ignores $S_A$ and maximizes $C^2/S_Q$, whereas energy subtraction instead minimizes $S_A$. Although $S_A$ and $S_Q$ can be expected to have completely different frequency distributions, and depending on the particular $F$, these two extremes are often good approximations, it is clearly a simplification. In this work a general optimization of Eq. (1) is instead considered.

If the low- and high-energy images are normalized with the expected number of counts from mean breast tissue, a combined image with zero mean can be formed according to

$$I(x,y) = w\frac{n_{lo}(x,y)}{\langle n_{lo}\rangle} + \frac{n_{hi}(x,y)}{\langle n_{hi}\rangle} - (w+1) \simeq w\ln\left[\frac{n_{lo}(x,y)}{\langle n_{lo}\rangle}\right] + \ln\left[\frac{n_{hi}(x,y)}{\langle n_{hi}\rangle}\right], \qquad (3)$$

where $w$ is a weight factor. The approximation is valid for $|n_\Omega - \langle n_\Omega\rangle| \ll 1$, i.e. small signal differences. Written this way, it is evident that a linear combination, which is the common form for energy weighting, is approximately equal to combination in the logarithmic domain, which is often used for energy subtraction. Energy weighting and energy subtraction can therefore be regarded special cases of a general image combination. Normalization with the expected number of counts in Eq. (3) is, however, a detour to make this point and to convey the derivations below. A consequence is that Eq. (2) does not make sense because $\langle I\rangle = 0$, but the product $\text{GNEQ} \times C^2$ is still valid. In the practical case, a combination of the non-normalized images is more handy, i.e.

$$I'(x,y) = w'n_{lo}(x,y) + n_{hi}(x,y) \quad\text{or}\quad I''(x,y) = w''\ln n_{lo}(x,y) + \ln n_{hi}(x,y). \qquad (4)$$

The image mean of $I$, $I'$, and $I''$ differ, but assuming small signal differences, the detectability indices calculated with all three image combinations are the same for

$$w' = \zeta_{hi}/\zeta_{lo} \times w \quad\text{and}\quad w'' = w. \qquad (5)$$

Note that $I'(w'=1)$ is a conventional non-energy-resolved absorption image.

If we assume no correlation between the energy bins, the quantum noise in the combined image is

$$S_Q(\omega) = \sum_\Omega \left.\frac{\partial I}{\partial n_\Omega}\right|_{\overline{n}_\Omega}^2 \times S_{Q\Omega}(\omega) \simeq \frac{1}{q_0}\left[\frac{w^2}{\zeta_{lo}} + \frac{1}{\zeta_{hi}}\right]. \qquad (6)$$

where $\Omega \in \{lo, hi\}$ denotes the detector energy bin, $q_0$ is the incident number of quanta, and $\zeta$ is the expected fraction of incident counts to be detected. The approximation in Eq. (6) is for spatially uncorrelated noise.

The anatomical noise in an x-ray image of breast tissue is caused by the variation in glandularity, which is transferred to the image through $I(g(x,y))$, with $g(x,y)$ being the glandular volume fraction as a function of spatial image coordinates $x$ and $y$. We therefore adopt the power spectrum of $g(x,y)$ as a glandularity NPS ($S_{Ag}(\omega)$), which is transferred to the image NPS ($S_A(\omega)$) according to

$$S_A(\omega) \simeq \left\langle\left|\frac{dI}{dg}\right|^2\right\rangle \times S_{Ag}(\omega)T^2(\omega) \simeq d_b^2[w\Delta\overline{\mu}_{\text{ag},lo} + \Delta\overline{\mu}_{\text{ag},hi}]^2 \times S_{Ag}(\omega)T^2(\omega), \qquad (7)$$

where $d_b$ is the breast thickness, $\Delta\overline{\mu}_{\text{ag},\Omega} \equiv \overline{\mu}_{a,\Omega} - \overline{\mu}_{g,\Omega}$ is the difference in effective linear attenuation between adipose and glandular tissue, and the angle brackets represent the expectation value over the glandularity range. The first approximation of Eq. (7) is for piecewise linearity of $I(g)$. The second approximation assumes linearity across the range of glandularities, image combination according to Eq. (3), and small signal differences.

Maximization of $\Delta s^2/S_Q$ and $1/S_A$ yields the optima for energy weighting and energy subtraction, respectively:

$$w^*_{s^2/S_Q} = \zeta_{lo}\Delta\overline{\mu}_{\text{bc},lo}/\zeta_{hi}\Delta\overline{\mu}_{\text{bc},hi}, \quad\text{and}\quad w^*_{1/S_A} = -\Delta\overline{\mu}_{\text{ag},hi}/\Delta\overline{\mu}_{\text{ag},lo}. \qquad (8)$$

$S_\mathrm{A}$ can in practice not be completely eliminated according to Eq. (8) because the latter is based on the linear approximation of $I(g)$ in Eq. (7), and a better estimate is to instead use the piecewise linear approximation in Eq. (7). Calculation of the expectation value, however, requires the probability density function ($\lambda_g$) according to

$$\left\langle \left|\frac{\mathrm{d}I}{\mathrm{d}g}\right|^2 \right\rangle = \int \left|\frac{\mathrm{d}I}{\mathrm{d}g}\right|^2 \times \lambda_g \mathrm{d}g \simeq \iint \left|\frac{\mathrm{d}I}{\mathrm{d}g}\right|^2 \times g(x,y) \mathrm{d}x \mathrm{d}y. \tag{9}$$

The approximation in Eq. (9) assumes a glandularity map ($g(x,y)$) to be a representative estimate of the density function.

## 2.2. Simulation of unenhanced spectral imaging

### 2.2.1. A model of the spectral imaging system

The previously developed model of the spectral imaging system[25, 26] was extended to include imaging of unenhanced tumors, microcalcifications and cysts in an anatomical background. Equation (1) was used as a figure of merit for optimization and for comparison to conventional absorption imaging. We assumed that the spectral image must come as a bonus on top of an optimal absorption image, which limited the choice of incident spectra and dose range. Compared to contrast-enhanced spectral imaging,[26] the split energy does not have to be matched to an absorption edge of the contrast agent, but can be chosen to minimize quantum noise. In addition, the sensitivity to a limited energy resolution can be expected to be lower in unenhanced imaging because of no discontinuities in the attenuation spectrum.

Energy resolved images were synthesized since clinical or phantom data was not available. The purpose of the images was twofold; to measure the noise in combined images for verification of Eqs. 6 and 7, and to visualize the result of image combination. A tungsten target x-ray tube with 0.5 mm aluminum filtration and 30 kVp acceleration voltage was assumed if not otherwise stated. The object was a 50 mm breast with 5 mm skin thickness and embedded lesions. Published x-ray spectra,[33] x-ray attenuation coefficients,[34] and dose coefficients[35] were used as input. Glandular structure was generated using the clustered lumpy background technique.[36] The structure was chosen to range over all glandularities with a 50% glandularity mean. Tumor x-ray attenuation was gathered from Johns and Yaffe,[16] calcium phosphate ($Ca_3P_2O_8$) microcalcifications and cysts consisting of 100% glandular tissue were assumed. The tumors and cysts were 20 mm thick and had diameters of 20 mm, 30 mm and 40 mm. Equal thickness means equal contrast, but the different diameters are affected differently by anatomical noise. Relatively large tumors were assumed to compensate for low detectability, but it has been shown that tumors larger than 20 mm constitute approximately 30% of all missed breast cancers.[37] The diameter of the calcifications was 100 $\mu$m. The objects were imaged at a dose of 1 mGy, and quantum noise was added with a fraction double-counted photons calculated by the detector model. We assumed that there were no dead channels in the detector.

### 2.2.2. Noise transfer

36 × 2 high- and low-energy images were generated with respectively pure anatomical and pure quantum noise, and the linear combination in Eq. (3) was used for image combination. An image size of 1536 × 1536 pixels (corresponding to ∼ 80 × 80 mm$^2$) was chosen. The NPS was measured in absorption and combined images, and the radial NPS was found by converting to polar coordinates and averaging over $2\pi$. For comparison, the NPS in the synthesized absorption images was fitted to analytical expressions as described below, and the NPS in combined images was calculated using these fits and the expressions in Section 2.1.3.

The NPS of anatomical backgrounds can be well described by an inverse power function, i.e.

$$S_\mathrm{A}(\omega) \simeq \alpha \omega^{-\beta}, \tag{10}$$

and this function was fitted to $S_\mathrm{A}$ in synthesized absorption images in a region that was virtually unaffected by window artifacts. The anatomical noise in combined images was calculated using the fit and Eq. (7), and could then be compared to measurements in combined synthesized images. To calculate the detectability index, a flat distribution of glandularities was assumed in Eq. (9), which can be expected to overestimate $S_\mathrm{A}$ since a Gaussian distribution is more probable, and predictions are therefore moderate. For comparison, $S_A$ was also calculated using the known glandularity map from the synthesized images, which is likely to give a better prediction.

Following the approximations in Eq. (7), the magnitude ($\alpha$) of $S_{Ag}(\omega)$ is affected by x-ray imaging and image combination, but the frequency dependence ($\beta$) is intact. It has been found previously that $\beta$ in images with large attenuation differences is in fact affected by the particular image combination,[27] which is likely to be at least partly caused by breakdown of the piecewise-linearity approximation in Eq. (7). For breast tissue, attenuation differences are relatively small and the approximation can be expected to hold better. The MTF of the imaging system, however, filters the image and therefore affects $\beta$, which is accounted for in Eq. (7).

Quantum noise with a fraction $\chi$ double-counted events is not completely flat in the frequency domain but follows

$$S_{Q\Omega}(u) = \bar{n}_\Omega \frac{1 + \chi_\Omega[1 + 2\cos(2\pi u/p)]}{1 + \chi_\Omega}, \qquad (11)$$

where $p$ is the pixel size and $u$ is the spatial frequency in the detector direction.[25] In the scan direction (spatial frequency $v$), $S_Q(v)$ was assumed flat because readouts are uncorrelated. $S_Q$ in combined images was calculated from Eq. (11) via Eq. (6). The latter is minimized for a bin count fraction

$$\xi_{lo} = 1 - \xi_{hi} = \frac{|w|}{1 + |w|}, \qquad (12)$$

which could determine a suitable split energy. If not otherwise stated, a count fraction close to 0.5 was, however, chosen in order to reduce complexity of the optimization.

### 2.2.3. Signal transfer and task function

The MTF was measured and fitted to an analytical function as previously described.[26,38] An assumption of equal MTF in both energy bins was adopted, although double counting degraded the resolution somewhat in the high-energy bin, which generally affects the GNEQ.[29] This simplifications was regarded justified because the difference in MTF between the bins was small, and the major differences between optimally combined images and conventional absorption images are in the region where anatomical noise dominates, i.e. at low spatial frequencies where $T(\omega) \simeq T_{lo}(\omega) \simeq T_{hi}(\omega) \simeq T(0) = 1$.

We used the designer nodule function, which was introduced by Burgess et al.,[31] to model the targets. For target radius $R$ and radial coordinate $r$, $s(r) \propto \text{rect}(\rho/2) \times (1 - \rho^2)^\nu$, where $\rho = r/R$. $\nu$ determines the shape of the function; $s$ is a projected sphere for $\nu = 0.5$, which we used to model microcalcifications, and approximates a tumor for $v = 1.5$. The Fourier transform of $s$ was used as task function in the model.

### 2.2.4. Images for visualization

For visualization of the optimal image combination, images with both quantum and anatomical noise were generated. Tumors and cysts, i.e. false positives, with profiles according to the designer nodule function were inserted. All images were filtered with the MTF.

Instead of the linear image combination that was used for system characterization, a polynomial-weighted logarithmic subtraction was introduced to gain better background subtraction than with a constant weight factor, i.e. $w = w(n_{lo}, n_{hi})$ in Eq. (3). The polynomial was trimmed to minimize the variance of a phantom with seven levels of glandularity. A second degree polynomial was found to provide a good-enough minimization. It can be noted that similar nonlinear techniques are employed for material-basis decomposition.[2,5] More efficient optimization schemes could be conceived, e.g. maximization of $C^2/S_A$.

Simulated images were low-pass filtered to reduce quantum noise. Equal filters were applied to both bins prior to forming the combined image. More advanced filtering methods have been shown to improve detectability considerably,[5,29] and also to influence the optimization because unequal filtering of the bins does not cancel in the GNEQ.

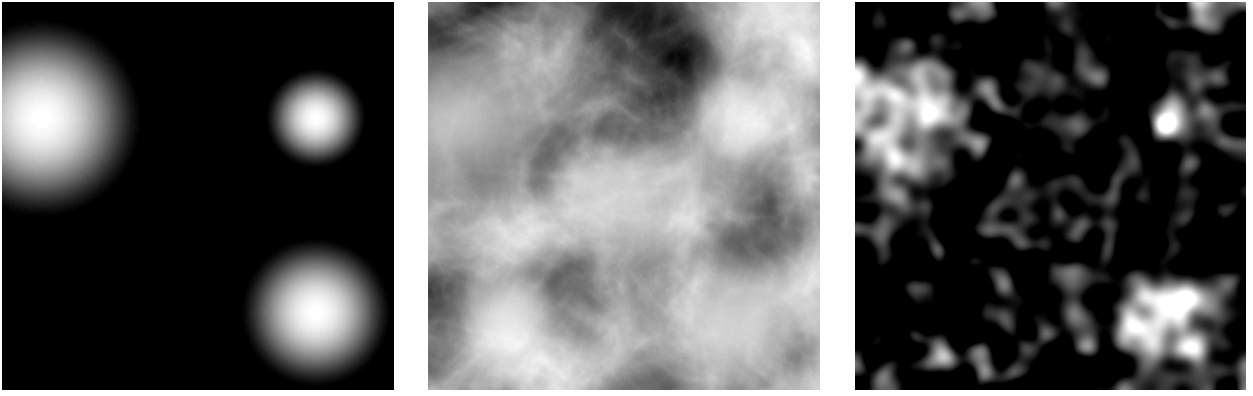

**Figure 2.** Synthetic images of three 20 mm thick tumors with diameters 20, 30, and 40 mm at an AGD of 1.0 mGy. **Left:** Tumor locations. **Center:** Absorption image with synthesized anatomical noise. **Right:** Optimally combined image, where all tumors are visible and all false positives are excluded.

## 3. RESULTS

### 3.1. Images for visualization

Synthesized images at 1 mGy of three 20 mm thick tumors with diameters 20, 30, and 40 mm embedded in a 50 mm breast are shown in Fig. 2. The left-hand image shows the tumors, which are hidden by anatomical and quantum noise in the center image. In addition, two cysts with equal size and attenuation as the tumors were added in the center and in the lower-left corner of the image. The polynomially combined image is shown to the right with all tumors visible and all false positives excluded. Anatomical noise was reduced at the cost of a lower contrast-to-quantum-noise ratio, and low-pass filtering with a 1.5 mm Gaussian kernel was therefore applied. It should be noted that filtering of the absorption image would reduce high-frequency noise, but not exclude the false positives. In addition, a narrow display window helped visualize the tumors in the combined image, but would not improve the absorption image. Smaller and thinner tumors than the ones imaged here were found hard to visualize because of quantum noise dominance and reduced contrast respectively.

### 3.2. Noise transfer

A logarithmic plot of the quantum and anatomical NPS in absorption images and images combined for maximum background subtraction is shown in Fig. 3 (Left). All 36 synthesized images were used in the calculations. $S_\mathrm{A}$ in the absorption image crossed $S_\mathrm{Q}$ at $\omega = 1.5$ mm$^{-1}$ with $\beta = 3.0$. Image combination with a constant weight factor reduced the quantum noise slightly, and reduced the anatomical noise more than three orders of magnitude. A polynomial weight factor reduced the anatomical noise another four orders of magnitude at equal quantum noise. Nevertheless, the following derivation of detectability index was based on the linear combination in Eq. (3) to ensure linearity, although the polynomial combination can be expected to perform better.

Figure 3 (Left) shows that $S_\mathrm{Q}$ calculated by Eq. (6) corresponded well to measurements in the synthesized images. $S_\mathrm{A}$ calculated with a flat glandularity distribution overestimated the noise, as expected, but a better prediction was offered by the distribution estimated from the glandularity map. In the calculation of detectability index below, the flat distribution was assumed so that predictions can be regarded moderate. The validity of the theoretical framework in Section 2.1.3 is further verified by Fig. 3 (Right), which plots the $S_\mathrm{A}$-$S_\mathrm{Q}$ crossing as a function of weight factor calculated with the glandularity-map distribution. The agreement between model and measurement was good at all weight factors. Figure 3 (Right) also plots $\alpha$ and $\beta$ as a function of weight factor. The assumption that $\beta$ is independent of image combination seems fair; there is only a slight dip at $w^*_{1/S_\mathrm{A}}$, whereas $\alpha$ varies over several orders of magnitude.

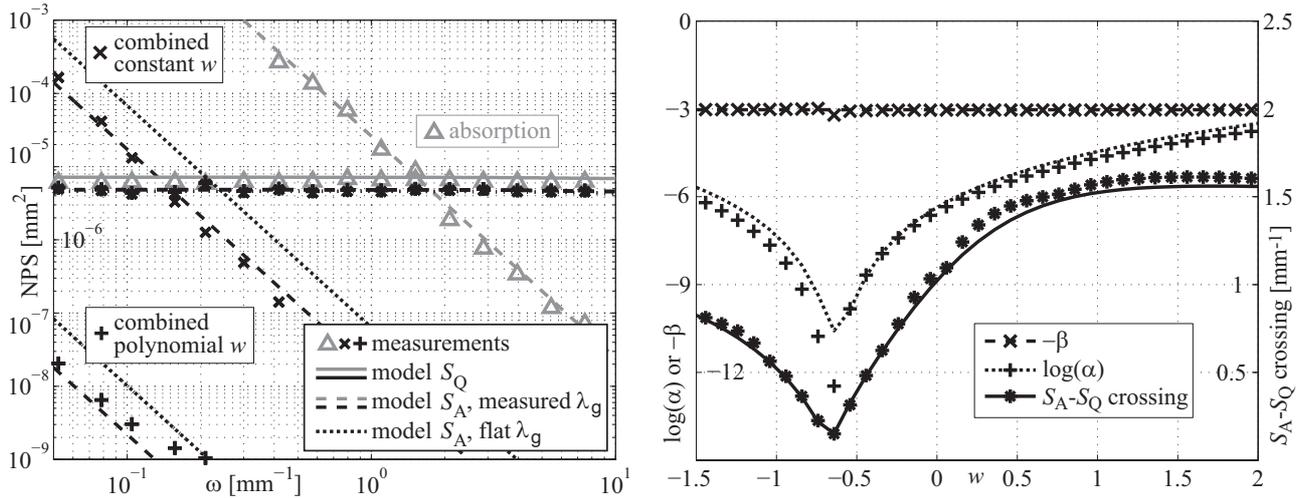

**Figure 3. Left:** Logarithmic plot of quantum and anatomical NPS ($S_Q$ and $S_A$) in synthesized absorption and combined images. Markers indicate measurements, and lines (denoted "model") show fits to measurements or predictions by the theoretical framework in Section 2.1.3. The results for two types of image combinations are presented: (1) a linear combination with a constant weight factor (constant $w$, as was used for calculating the detectability index), and (2) combination with a polynomial weight factor (polynomial $w$, as was used for the righthand image in Fig. 2). For each image combination, calculations of $S_A$ are presented for (1) a flat glandularity distribution (flat $\lambda_g$) and (2) the distribution measured from the synthesized glandularity map (measured $\lambda_g$). **Right:** The logarithm of the magnitude ($\log(\alpha)$) and the exponent ($-\beta$) of $S_A$ in a combined image are plotted against the left axis as a function of a constant weight factor ($w$). The crossing between anatomical and quantum noise ($S_A$-$S_Q$) for the same image combination is plotted against the right axis. Markers indicate measurements in the synthesized images and lines show predictions by the theoretical framework in Section 2.1.3.

### 3.3. Generalized NEQ and detectability index

The detectability index was calculated for the case shown in Fig. 2, and the NPS used for the calculations was therefore measured in this image alone. The measurement provided a higher NPS than for all 36 images combined (as considered in Fig. 3) with an $S_A$-$S_Q$ crossing at $\omega = 2.3$ mm$^{-1}$ and $\beta = 3.1$.

Detectability index as a function of weight factor is plotted in Fig. 4 (Left), and for three different cases: (1) a 20 mm tumor in anatomical and quantum noise (similar to Fig. 2), (2) a 100 $\mu$m microcalcification in anatomical and quantum noise, and (3) a tumor on a flat background with only quantum noise. Note that in Fig. 4 (Left), all detectability indices are normalized to the absorption image so that the plot shows the benefit of spectral imaging. In addition, positive and negative weight factors are reported as $w'$ and $w''$ according to Eq. (5). Put together, this means that the absorption images of all three cases are located at (1,1).

For the tumor in anatomical noise, optimal combination was found to be close, but clearly not identical, to energy subtraction according to Eq. (8). Energy weighting was suboptimal because higher weighting of the low-energy photons also increased the anatomical noise. For the small microcalcification, however, energy subtraction was suboptimal whereas energy weighting provided a minute improvement. A similar result was found for the tumor on a uniform background, and the optimal weight factors for these targets almost coincided. The optimal weight factor for energy weighting, and also the optimal energy, thus seems fairly independent of lesion type, which is in accordance with previous studies.[18,39]

Figure 4 (Right) shows the different parts of Eq. (1) at optimal image combination for the 20-mm tumor and the 100-$\mu$m microcalcification in anatomical noise. The tumor contrast-to-noise ratio (GNEQ$(\omega) \times C^2$) of the combined image was relatively low, and benefit over the absorption image was found only at frequencies below $\sim 0.5$ mm$^{-1}$, where, however, the tumor task function ($F^2(\omega) \times \omega$) was located. The microcalcification task function, on the other hand, increased with spatial frequency due to the two-dimensional integration, and

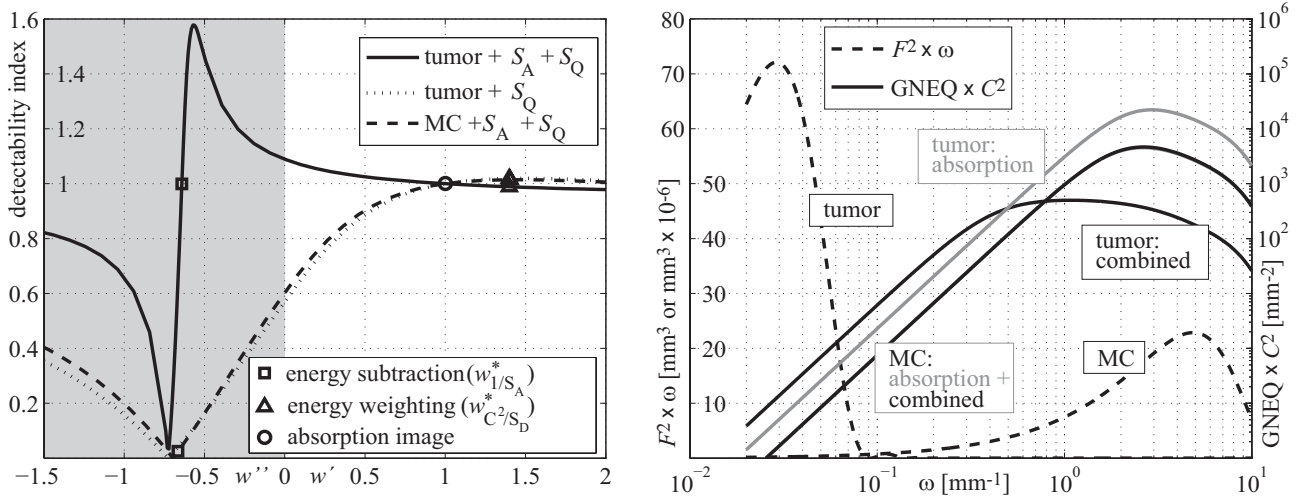

**Figure 4. Left:** Detectability as a function of weight factor ($w$) normalized so that a conventional absorption image is located at (1,1). Three cases are presented: (1) a 20-mm tumor in anatomical and quantum noise (tumor + $S_A$ + $S_Q$), (2) the tumor on a homogenous background with quantum noise only (tumor + $S_Q$), and (3) a 100-$\mu$m microcalcification in both types of noise (MC + $S_A$ + $S_Q$). Markers indicate weight factors for the absorption image and for optimal image combinations according to Eq. (8). **Right:** The task functions ($F^2 \times \omega$) for a 20-mm tumor and a 100-$\mu$m microcalcification (MC) are plotted against the left axis. The microcalcification task function is multiplied with $10^6$ to make it visible in the figure. Logarithmic plots of the contrast-to-noise ratio squared (GNEQ $\times C^2$) for absorption and optimally combined images of these two targets are shown against the right axis. The microcalcification plots virtually coincide. A combination of $F^2 \times \omega$ and GNEQ $\times C^2$ illustrates integration to detectability index in Eq. (1).

was hence virtually unaffected by the anatomical noise. Energy subtraction was therefore suboptimal and a small benefit was provided by energy weighting, which increased the contrast-to-noise ratio almost equally for all spatial frequencies.

A split energy of 21 keV provided a bin count fraction of approximately $\xi_{lo} = \xi_{hi} = 0.5$, and was used for all of the above cases. A scan of split energies at the optimal weight factor for tumor imaging (-0.57) revealed that the optimum for this case was 18.0 keV, which, however, improved the detectability on the order of 1%, and the spectrum can hence be safely split at the center. It can be noted that 18.0 keV yields $\xi_{lo} = 0.29$, which is close to $\xi_{lo} = 0.36$ as predicted by Eq. (12).

Table 1 presents detectability indices for absorption and optimally combined images. The first row concerns the cases considered above: a 20-mm tumor and a 100-$\mu$m microcalcification in quantum and anatomical noise imaged with 30 kV and the experimental detector. Detectability of the tumor can be improved 50% by optimal combination, which seems reasonable when comparing to Fig. 2. Optimal combination for imaging the microcalcification yielded only a slight improvement on the order of 1%.

The subsequent rows show two cases of relatively straightforward system changes that might influence detectability: a change in beam quality (higher acceleration voltage), and an optimized detector with electronic noise, channel-to-channel threshold spread, dead time, and AC leakage reduced by a factor of two. Neither beam quality nor improved detector performance was found to influence the result materially for any of the cases. For combined images, a harder spectrum reduced detectability of both tumors and microcalcifications slightly, whereas optimized detector performance provided a minute improvement. For absorption images, the harder spectrum reduced detectability of microcalcifications, but improved the result for tumors because the reduction in background contrast, i.e. anatomical noise, was larger than the reduction in tumor contrast.

**Table 1.** Detectability index ($d'$) for optimally combined and conventional absorption images of tumors and microcalcifications (MCs). Optimization was done with acceleration voltage and the experimental compared to an optimized detector.

|   | detector | acc. voltage [kV] | Al filter [mm] | $E_{\text{split}}$ [keV] | $d'$ **combined / absorption** 20 mm tumor | 0.1 mm MC |
|---|---|---|---|---|---|---|
| 1. | experimental | 30 | 0.5 | 21 | **6.32 / 4.20** | **3.10 / 3.06** |
| 2. | experimental | 40 | 0.5 | 25 | **6.06 / 4.34** | **2.86 / 2.74** |
| 3. | optimized | 30 | 0.5 | 21 | **6.36 / 4.00** | **3.16 / 3.10** |

## 4. DISCUSSION

It can be questioned if the level of the anatomical noise in the synthesized images corresponds to real breast tissue. The exponent ($\beta$) seems to be in the reasonable range; values between three and four have been published for breast tissue.[31, 36] The magnitude ($\alpha$), on the other hand, was difficult to validate. Burgess reported that the anatomical noise in digitized mammograms dominated below $\sim 1$ mm$^{-1}$,[31] which is slightly lower than the $S_A$-$S_Q$ crossings of the synthesized images and indicates that we used a noise magnitude that was higher than average but not totally arbitrary. A measurement of the NPS in clinical images with the multi-slit system would be required to fully settle this issue.

Another uncertainty concerns the agreement between the model and real observers. The ideal observer that was used in this study represents the upper limit of observer performance, and other models, such as the non-prewhitening observer has been shown to provide better agreement with human observers in some cases.[30] In addition, it has been shown that the noise in mammograms is not completely random, as was assumed here, but has a deterministic component that the radiologist after training may be able to see through.[32] Finally, factors other than anatomical noise, such as variations in thickness or anatomy, might play a big role in the practical case. Observer studies in realistic anatomical backgrounds and clinical studies are needed to fully investigate these effects.

There are several potential improvements to the technology, which warrant further study. These include optimization of spatial filtering for noise reduction,[29] optimization of incident spectrum,[15, 40] and nonlinear image combination.[1, 2, 5]

The results presented here for unenhanced spectral imaging also have implications to non-energy-resolved beam quality optimization. We saw in Table 1 that detectability for microcalcifications in the absorption image was reduced by a harder spectrum, which is in line with common optimization that only includes quantum noise. The detectability for tumors, which are more affected by anatomical noise, however, rose with the harder spectrum. This is further illustrated in Fig. 4 (Left), where excluding the low-energy image altogether ($w = 0$) resulted in higher detectability than was found for the absorption image. In addition, comparing Eqs. (6) and (7), we see that $S_A$ as opposed to $S_Q$ is independent of dose. Increasing the dose therefore does not improve the GNEQ if anatomical noise dominates, contrary to the standard NEQ.

## 5. CONCLUSIONS

Unenhanced spectral imaging has great potential because it comes as a bonus to the conventional non-energy-resolved absorption image at screening; there is no additional radiation dose to the patient and no need to inject contrast medium. We have used a previously developed theoretical framework and system model to characterize the performance of a photon-counting spectral imaging system with two energy bins for unenhanced spectral mammography. The model calculated a task-dependent ideal-observer detectability index via the generalized NEQ (GNEQ), which includes quantum and anatomical noise. This figure of merit was used to find an optimal combination of the energy-resolved images, to compare optimally combined images with absorption images, and to investigate the effect of system optimization. In addition, synthesized images with quantum and anatomical noise were generated with the system model and used to verify the theoretical framework and to illustrate the technique.

Optimal combination for imaging of large unenhanced tumors in the presence of anatomical noise provided a 50% improvement in detectability compared to absorption imaging. The image combination corresponded closely, but not exactly, to minimization of the anatomical noise, i.e. the energy subtraction scheme. Higher weighting of the more-information-dense photons, referred to as energy weighting, deteriorated detectability for this task. For small microcalcifications or tumors on uniform backgrounds, however, the situation was reversed; energy subtraction was clearly suboptimal whereas energy weighting provided a small benefit on the order of 1%. The performance was largely independent of beam quality, detector energy resolution, and bin count fraction, which simplifies optimization in the practical case. Several potential improvements to the technique warrant further study, including spatial filtering and nonlinear image combination.

Optimal image combination and the benefit of spectral imaging depended to a large extent on the anatomical noise and imaging task. This may have implications also on optimization of non-energy resolved imaging, where it is common practice to consider quantum noise alone.

# REFERENCES

1. Alvarez, R. and Macovski, A., "Energy-selective reconstructions in x-ray computerized tomography," *Phys. Med. Biol.* **21**, 733–744 (1976).
2. Lehmann, L. A., Alvarez, R. E., Macovski, A., Brody, W. R., Pelc, N. J., Riederer, S. J., and Hall, A. L., "Generalized image combinations in dual KVP digital radiography," *Med. Phys.* **8**(5), 659–667 (1981).
3. Baldelli, P., Bravin, A., Maggio, C. D., Gennaro, G., Sarnelli, A., Taibi, A., and Gambaccini, M., "Evaluation of the minimum iodine concentration for contrast-enhanced subtraction mammography," *Phys. Med. Biol.* **51**(17), 4233–51 (2006).
4. Lewin, J., Isaacs, P., Vance, V., and Larke, F., "Dual-energy contrast-enhanced digital subtraction mammography: Feasibility," *Radiology* **229**, 261–268 (2003).
5. Johns, P. and Yaffe, M., "Theoretical optimization of dual-energy x-ray imaging with application to mammography," *Med. Phys.* **12**, 289–296 (1985).
6. Johns, P., Drost, D., Yaffe, M., and Fenster, A., "Dual-energy mammography: initial experimental results," *Med. Phys.* **12**, 297–304 (1985).
7. Bliznakova, K., Kolitsi, Z., and Pallikarakis, N., "Dual-energy mammography: simulation studies," *Phys. Med. Biol.* **51**(18), 4497–4515 (2006).
8. Boone, J., Shaber, G., and Tcotzky, M., "Dual energy mammography: a detector analysis," *Med. Phys.* **17**, 665–675 (1990).
9. Chakraborty, D. P. and Barnes, G. T., "An energy sensitive cassette for dual-energy mammography," *Med. Phys.* **16**, 7–13 (1989).
10. Kappadath, S. C. and Shaw, C. C., "Qunatitative evaluation of dual-energy digital mammography for calcification imaging," *Phys. Med. Biol.* **53**(19), 5421–5443 (2008).
11. Brettle, D. S. and Cowen, A. R., "Dual-energy digital mammography utilizing stimulated phosphor computed radiography," *Phys. Med. Biol.* **39**(11), 1989–2004 (1994).
12. Asaga, T., Chiyasu, S., Mastuda, S., Mastuura, H., Kato, H., Ishida, M., and Komaki, T., "Breast imaging: dual-energy projection radiography with digital radiography," *Radiology* **164**, 869–870 (1987).
13. Masuzawa, T. A. C., Yoshida, A., and Mastuura, H., "Dual-energy subtraction mammography," *J. Digit. Imaging* **8**(1), 70–73 (1995).
14. Taibi, A., Fabbri, S., Baldelli, P., di Maggio, C., Gennaro, G., Marziani, M., Tuffanelli, A., and Gambaccini, M., "Dual-energy imaging in full-field digital mammography: a phantom study," *Phys. Med. Biol.* **48**, 1945–1956 (2003).
15. Avila, C., Lopez, J., Sanabria, J. C., Baldazzi, G., Bollini, D., Gombia, M., Cabal, A. E., Ceballos, C., Garcia, A. D., Gambaccini, M., Taibi, A., Sarnelli, A., Tuffanelli, A., Giubellino, P., Marzari-Chiesa, A., Prino, F., Tomassi, E., Grybos, P., Idzik, M., Swientek, K., Wiacek, P., no, L. M. M., Ramello, L., and Sitta, M., "Contrast cancellation technique applied to digital x-ray imaging using silicon strip detectors," *Med. Phys.* **32**(12), 3755–3766 (2005).
16. Johns, P. C. and Yaffe, M. J., "X-ray characterisation of normal and neoplastic breast tissues," *Phys. Med. Biol.* **32**(6), 675–695 (1987).

17. Tapiovaara, M. and Wagner, R., "SNR and DQE analysis of broad spectrum x-ray imaging," *Phys. Med. Biol.* **30**, 519–529 (1985).
18. Cahn, R., Cederström, B., Danielsson, M., Hall, A., Lundqvist, M., and Nygren, D., "Detective quantum efficiency dependence on x-ray energy weighting in mammography," *Med. Phys.* **26**(12), 2680–3 (1999).
19. Bornefalk, H., Lewin, J. M., Danielsson, M., and Lundqvist, M., "Single-shot dual-energy subtraction mammography with electronic spectrum splitting: Feasibility," *Eur. J. Radiol.* **60**, 275–278 (2006).
20. Bornefalk, H., Hemmendorff, M., and Hjärn, T., "Contrast-enhanced dual-energy mammography using a scanned multislit system: evaluation of a differential beam filtering technique," *J. Electron. Imaging* **16**(2) (2007).
21. "The HighReX Project (High Resolution X-ray imaging)." online: http://www.highrex.eu.
22. Lundqvist, M., *Silicon strip detectors for scanned multi-slit x-ray imaging*, PhD thesis, Royal Institute of Technology (KTH), Stockholm (2003).
23. Åslund, M., Cederström, B., Lundqvist, M., and Danielsson, M., "Physical characterization of a scanning photon counting digital mammography system based on Si-strip detectors," *Med. Phys.* **34**(6), 1918–1925 (2007).
24. Åslund, M., Cederström, B., Lundqvist, M., and Danielsson, M., "Scatter rejection in multi-slit digital mammography," *Med. Phys.* **33**, 933–940 (2006).
25. Fredenberg, E., Lundqvist, M., Cederström, B., Åslund, M., and Danielsson, M., "Energy resolution of a photon-counting silicon strip detector," *Nucl. Instr. and Meth. A* **613**(1), 156–162 (2010).
26. Fredenberg, E., Hemmendorff, M., Cederström, B., Åslund, M., and Danielsson, M., "Contrast-enhanced spectral mammography with a photon-counting detector," *Med. Phys.* (2009). Submitted for publication.
27. Richard, S., Siewerdsen, J. H., Jaffray, D. A., Moseley, D. J., and Bakhtiar, B., "Generalized DQE analysis of radiographic and dual-energy imaging using flat-panel detectors," *Med. Phys.* **32**(5), 1397–1413 (2005).
28. Richard, S. and Siewerdsen, J. H., "Optimization of dual-energy imaging systems using generalized NEQ and imaging task," *Med. Phys.* **34**(1), 127–139 (2007).
29. Richard, S. and Siewerdsen, J. H., "Cascaded systems analysis of noise reduction algorithms in dual-energy imaging," *Med. Phys.* **35**(2), 586–601 (2008).
30. Richard, S. and Siewerdsen, J. H., "Comparison of model and human observer performance for detection and discrimination tasks using dual-energy x-ray images," *Med. Phys.* **35**, 5043–5053 (2008).
31. Burgess, A. E., Jacobson, F. L., and Judy, P. F., "Human observer detection experiments with mammograms and power-law noise," *Med. Phys.* **28**(4), 419–437 (2001).
32. Bochud, F., Valley, J., Verdun, F., Hessler, C., and Schnyder, P., "Estimation of the noisy component of anatomical backgrounds," *Med. Phys.* **26**(7), 1365–1370 (1999).
33. Boone, J., Fewell, T., and Jennings, R., "Molybdenum, rhodium, and tungsten anode spectral models using interpolating polynomials with application to mammography," *Med. Phys.* **24**(12), 1863–74 (1997).
34. Berger, M. J., Hubbell, J. H., Seltzer, S. M., Coursey, J. S., and Zucker, D. S., "XCOM: Photon Cross Section Database." http://physics.nist.gov/xcom. National Institute of Standards and Technology, Gaithersburg, MD (2005).
35. Boone, J., "Glandular breast dose for monoenergetic and high-energy x-ray beams: Monte Carlo assessment," *Radiology* **203**, 23–37 (1999).
36. Bochud, F., Abbey, C., and Eckstein, M., "Statistical texture synthesis of mammographic images with clustered lumpy backgrounds," *Opt. Express* **4**(1), 33–43 (1999).
37. Reintgen, D., Berman, C., Cox, C., Baekey, P., Nicosia, S., Greenberg, H., Bush, C., Lyman, G. H., and Clark, R. A., "The anatomy of missed breast cancers," *Surgical Oncology* **2**(1), 65 – 75 (1993).
38. Fredenberg, E., *Spectral Mammography with X-Ray Optics and a Photon-Counting Detector*, PhD thesis, Royal Institute of Technology (KTH), Stockholm (2009).
39. Fahrig, R. and Yaffe, M. J., "Optimization of spectral shape in digital mammography: dependence on anode material, breast thickness, and lesion type," *Med Phys* **21**(9), 1473–81 (1994).
40. Fredenberg, E., Cederström, B., Lundqvist, M., Ribbing, C., Åslund, M., Diekmann, F., Nishikawa, R., and Danielsson, M., "Contrast-enhanced dual-energy subtraction imaging using electronic spectrum-splitting and multi-prism x-ray lenses," in [*Proc. SPIE, Physics of Medical Imaging*], Hsieh, J. and Samei, E., eds., **6913** (2008).